\documentclass[12pt]{iopart}
 \usepackage{epsfig}

\begin{document}

\title {Wonderful life at weak Coulomb interaction:
increasing of superconducting/superfluid
transition temperature by disorder}

\author{V.E.Kravtsov}

\address{Abdus Salam ICTP, P.O.Box 586, 34100 Trieste, Italy}
\ead{kravtsov@ictp.it} \maketitle
\begin{abstract}
We have shown that in systems where the Coulomb interaction is
strongly suppressed, the superconducting transition temperature may
be enhanced by disorder close to the Anderson localization
transition. This phenomenon is based on the enhancement by disorder
of the matrix element of attraction in the Cooper channel. For
enhancement to take place one needs (i) strong disorder which makes
the single-particle wave functions strongly inhomogeneous in space
and (ii) strong correlation of the patterns of inhomogeneity for
different wavefunctions. One case where such correlation is well
known is the system close to the Anderson transition. We review the
notion of multifractality of wavefunctions in this region and show
how the enhancement of $T_{c}$ arises out of the multifractal
correlations.
\end{abstract}

\maketitle

\section{Introduction}
The statement which is known under the name of the "Anderson
theorem" and belongs to  P.W.Anderson \cite{Anderson-theorem} and
Abrikosov and Gorkov \cite{AGD}, reads that the non-magnetic
disorder does not change the superconducting transition temperature
$T_{c}$. As a matter of fact this statement has a status of a
theorem based on the normalization and completeness of the set of
single-particle wave functions only if the paring amplitude $\Delta$
does not have any variations in space or the latter can be averaged
independently of the fluctuations of the single-particle wave
functions. This is not true for sufficiently strong disorder where
the joint effect of disorder and Coulomb interaction leads to
suppression of $T_{c}$. This effect studied in detail by Finkelstein
\cite{Fink} was actually predicted earlier \cite{MayFuk} as the
leading correction to the mean field transition temperature in 2D
disordered superconductors:
\begin{equation}
\label{MayFuk} \frac{\delta T_{c}}{T_{c}}=-\frac{\lambda_{{\rm
eff}}}{ 3g}\,\ln^{3}\left(
\frac{\omega_{0}}{T_{c}}\right),\;\;\;\;g=2\pi^{2}\nu_{0}D.
\end{equation}
where $\lambda_{{\rm eff}}=\nu_{0}V$ is the dimensionless constant
of local interaction $V_{q}=V$, $g$ is the dimensionless
conductance, $\nu_{0}$ is the DoS at the Fermi level, $D$ is the
diffusion constant and $\omega_{0}$ is the Debye frequency.

For the screened Coulomb interaction in the universal limit
$\lambda_{{\rm eff}}=1$ and Eq.(\ref{MayFuk}) corresponds to
suppression of $T_{c}$ by disorder.
\begin{equation}
\label{MayFuk-Coulomb} \left(\frac{\delta T_{c}}{T_{c}}\right)_{{\rm
Coulomb}}=-\frac{ 1}{3g}\,\ln^{3}\left(
\frac{\omega_{0}}{T_{c}}\right),
\end{equation}
 However, if one assumes that the
Coulomb interaction is absent (or strongly suppressed) and the only
interaction that remains is the attraction in the Cooper channel
characterizing by the small dimensionless constant $\lambda_{{\rm
eff}}=-\lambda$, the same Eq.(\ref{MayFuk}) together with the BCS
relation $\lambda^{-1}=\ln(\omega_{0}/T_{c})$ results in:
\begin{equation}
\label{MayFuk-attrac} \left(\frac{\delta T_{c}}{T_{c}}\right)_{{\rm
Cooper}}=+\frac{ 1}{3g}\,\ln^{2}\left(
\frac{\omega_{0}}{T_{c}}\right),
\end{equation}
Thus in the absence of Coulomb interaction one obtains the
enhancement (w.r.t. the BCS result) of $T_{c}$ by disorder by the
same token as the suppression in the universal limit of the screened
Coulomb interaction known as the Finkelstein effect \cite{Fink}.

It was not understood in the eighties that the origin of both
suppression and enhancement is the weak multifractality of
single-particle wave functions in 2D metals. The main goal of this
paper is to make this connection physically transparent.
\section{What is the multifractality?}
While the earlier period of Anderson localization was mostly devoted
to study of the scaling behavior of the localization or correlation
length as a function of proximity to the mobility edge $E-E_{c}$ or
to critical disorder $W-W_{c}$, the notion of multifractality of
random critical single-particle wave function $\psi_{n}({\bf r})$
was introduced by Wegner \cite{Wegner} almost immediately after the
formulation of the scaling theory of Anderson transition. The
definition was given in terms of the inverse participation ratio and
its moments:
\begin{equation}
\label{def-frac} \int d^{d}{\bf r}\,\langle |\psi_{n}({\bf r})|^{2q}
\rangle\propto L^{-(q-1)\,d_{q}},
\end{equation}
where $\langle...\rangle$ denotes disorder average, $L$ is the size
of the system and $d_{q}$ are the {\it fractal dimensions}.
Eq.(\ref{def-frac}) covers all the three possible phases. The metal
phase is characterized by the wave functions which occupy all the
available volume with $|\psi_{n}({\bf r})|^{2}\sim L^{-d}$ by
normalization. In this case one immediately concludes from
Eq.(\ref{def-frac}) that all $d_{q}$ are equal to the space
dimensionality $d$. In the Anderson insulator $|\psi_{n}({\bf
r})|^{2}$ is of the order of $|\psi_{n}({\bf r})|^{2}\sim \xi^{-d}$
with the probability $(\xi/L)^{d}$ for a point ${\bf r}$ to fall
inside the localization volume $\xi^{d}$, and is very small
otherwise with the overwhelming probability $1- (\xi/L)^{d}$. Thus
the leading term in Eq.(\ref{def-frac}) at large $L$ is of the order
of $L^{d}\xi^{-qd}(\xi/L)^{d}\sim \xi^{-(q-1)d}$ and hence it is
independent of $L$, which formally corresponds to all $d_{q}=0$. In
the critical region close to the Anderson transition the set of
critical exponents $d_{q}$ is the an important and non-trivial
characteristic of this transition, in addition to the well known
exponent $\nu$ of the localization length $\xi\propto
|W-W_{c}|^{-\nu}$.

Another way to characterize multifractality is to define the {\it
spectrum of fractal dimensions} $f(\alpha)$ so that the volume in
space $V_{\alpha}$ where the amplitude $|\psi_{n}({\bf r})|^{2}$
scales like $L^{-\alpha}$ is  $V_{\alpha}\propto L^{f(\alpha)}$.
This approach explicitly takes into account the main feature of
multifractality: the hierarchy of regions in space with different
scaling of $|\psi|^{2}$ with the system size, each region being a
fractal of the Hausdorf dimension $f(\alpha)$. With this definition
we have for the integral of Eq.(\ref{def-frac}):
\begin{equation}
\label{Legendre-int} L^{-(q-1)d_{q}} \propto \int_{0}^{\infty}
d\alpha\, L^{-q\alpha+f(\alpha)}.
\end{equation}
By doing this integral in the saddle-point approximation valid at
large $\ln L$ one concludes that the quantity
$\tau_{q}=(q-1)\,d_{q}$ and $f(\alpha)$ are related by the Legendre
transformation:
\begin{equation}
\label{Legendre} \tau_{q}=\alpha\,q -
f(\alpha),\;\;\;\;\frac{df(\alpha)}{d\alpha}=q.
\end{equation}
The validity of scaling Eq.(\ref{def-frac}) at the Anderson
transition point was demonstrated in numerous numerical studies and
analytically by $\epsilon=d-2$ expansion \cite{Wegner} of the
nonlinear sigma-model (weak multifractality $d-d_{q}<<d$) or by the
virial expansion method \cite{KravEvt} (strong multifractality
$d_{q}<<d$).

For further discussion it is very important to know the correlation
of {\it two} single particle wavefunctions with the energies
$E_{n}=E$ and $E_{m}=E+\omega$ which determines the matrix element
of the local interaction. At the Anderson transition point this
matrix element appears to be a power-law function of the energy
difference $\omega$ as was conjectured by Chalker \cite{Chalker},
checked by direct diagonalization of the lattice Anderson models and
critical random matrix ensembles \cite{CueKrav} and finally proven
analytically for the critical random matrix ensemble \cite{KOY}:
\begin{equation}
\label{Chalk}M_{\omega}= L^{d}\int d^{d}{\bf r}\, \langle
|\psi_{E}({\bf r})|^{2}\,|\psi_{E+\omega}({\bf r})|^{2}\rangle \sim
  \left(\frac{E_{0}}{\omega}\right)^{\mu},
\end{equation}
where the exponent $\mu$ is closely related \cite{Chalker} with the
fractal dimension $d_{2}$:
\begin{equation}
\label{mu} \mu=1-\frac{d_{2}}{d}.
\end{equation}

The scaling Eq.(\ref{Chalk}),(\ref{mu}) in the frequency domain
reflects the critical nature of single-particle states near the
Anderson transition. However, it is valid well outside the mobility
edge.
\begin{figure}[h]
\includegraphics[width=10cm, height=8cm,angle=0]{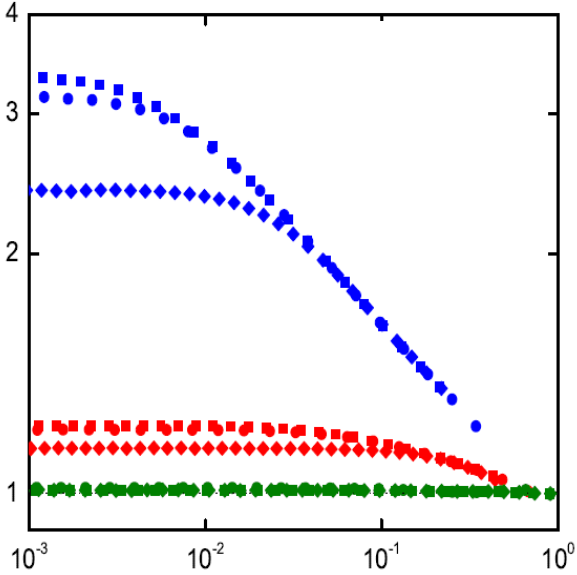}
\caption{(color online) The correlation function Eq.(\ref{Chalk}) as
a function of $\omega$ from direct diagonalization of the 3D
Anderson model with critical disorder W=16.5 for disorder strength
W=2 (green), W=5 (red) and W=10 (blue), and different system sizes
L=8(diamond),L=16 (circles),L=20 (squares). At $W=10$ and $L=16,20$
the critical power-law is well seen for
$(\nu_{0}\xi^{d})^{-1}=\delta_{\xi}<\omega<E_{0}$.}
\label{corr-metal}
\end{figure}
In Fig.1 it is shown the result of  calculation of the correlation
function Eq.(\ref{Chalk}) by direct diagonalization of the 3D
Anderson model in the metallic phase with sub-critical disorder. One
can see that even relatively far from the critical disorder
($W_{c}=16.5$), at $W=10$, the power-law critical behavior is well
seen. It is saturated when the  $\omega$-dependent "resolution
length" $L_{\omega}=(\nu_{0}\omega)^{-1/d}$ becomes larger than
either the correlation length $\xi\propto (W_{c}-W)^{-\nu}$ or the
system size $L$. Even at disorder as small as $W=5$ the enhancement
of $M_{\omega}$ given by the factor $(E_{0}/\omega)^{\mu}$ in
Eq.(\ref{Chalk}) is quite significant. Only at $W=2$ when
$\xi<\ell=(\nu_{0}E_{0})^{\frac{1}{d}}$ it disappears, and we reach
the limit $L^{d}M_{\omega}=1$ expected for the extended single
particle wavefunctions which occupy all the available space.

Such a behavior of the correlation function Eq.(\ref{Chalk})
suggests a stricture of the typical single-particle wavefunction at
the sub-critical disorder which is sketched in Fig.2.
 \begin{figure}[h]
\includegraphics[width=10cm, height=8cm,angle=0]{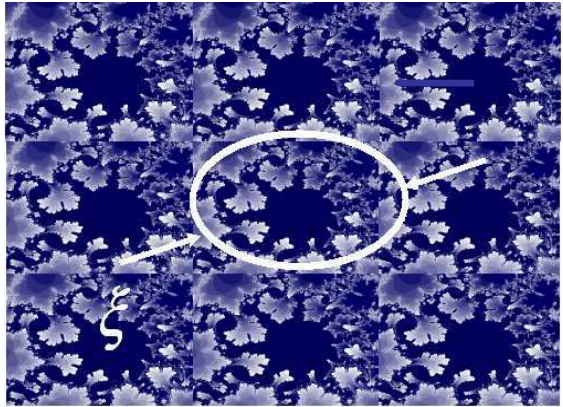}
\caption{(color online) The cartoon of the amplitude landscape of
the sub-critical single-particle wavefunction. The correlation
length $\xi\propto (W_{c}-W)^{-\nu}$ sets the typical size of the
"fractal region" where the wavefunction correlations are the same as
in the critical state. If the $\omega$-dependent "resolution length"
$L_{\omega}=(\nu_{0}\omega)^{-\frac{1}{d}}$ is smaller than $\xi$
the critical power law in Eq.(\ref{Chalk}) is well seen. For larger
resolution length one observes a saturation in the
$\omega$-dependence which is typical of metal. However, the fact
that there are many peaks (dark regions on the plot)in the
wavefunction landscape where the
 amplitude is pretty high, leads to the enhanced saturation level.
The exception is the case of a very weak disorder when the
correlation length $\xi\ll \ell$ is smaller than the size
$\ell=(\nu_{0}E_{0})^{-\frac{1}{d}}$ of the "pixel" which is only
slightly larger than the atomic scale. Only in this limit the
Anderson theorem may apply. The similar cartoon holds for the
localized states at super-critical disorder. In this case the
amplitude is exponentially small outside the localization radius
$\xi$ but has the same  fractal structure inside this radius. That
allows to speak about the "multifractal metal" and "multifractal
insulator" \cite{CueKrav}. } \label{cartoon-metal}
\end{figure}
It consists of the "fractal regions" of the size of the "correlation
radius" $\xi$ appearing in the scaling theory of localization.
Inside such regions all the correlations are identical to those in
the critical wavefunction. As the energy difference $\omega$
decreases the "resolution length"
$L_{\omega}=(\nu_{0}\omega)^{-\frac{1}{d}}$ increases and at
$\omega<\delta_{\xi}=(\nu_{0}\xi^{d})^{-1}$ it exceeds the
correlation length $\xi$. Only for such $\omega$ the correlation
function Eq.(\ref{Chalk}) may show that the state is not critical
but extended. Thus $\omega=\delta_{\xi}$ marks the onset of
saturation in Fig.1 for $L\gg \xi$.

It is important that at large $\omega=E_{0}$   all curves in Fig.1
merge, in agreement with Eq.(\ref{Chalk}). This means that for
$\omega<E_{0}$ the matrix element $M_{\omega}$ is {\it enhanced}
compared to the ideal case of totally extended wavefunctions which
occupy all the available space without peaks and holes at a scale
much larger than the Fermi wavelength. In order to better understand
the origin of this enhancement we emphasize that it is caused by the
{\it inhomogeneity} of the amplitude $|\psi({\bf r})|^{2}$ in space.
The presence of "holes" with very small amplitude in certain regions
implies appearance of peaks in other regions, as the integral of the
amplitude over entire system of the size $L$ is equal to 1 by
normalization. The key point is that the positions of the peaks in
two different wave functions are highly correlated for the critical
and slightly off-critical states. Contrary to naive expectation,
even at large energy difference $\omega$ which is only slightly
smaller than the {\it upper energy scale for fractality} $E_{0}$
(for the 3D Anderson model it is approximately 1/3 of the
bandwidth), the correlation is pretty high. This leads to the very
slow decay of the matrix element $M_{\omega}\sim \omega^{-\mu}$
($\mu<1$) with increasing $\omega$ \cite{Kmut,MirFyod}. Note that
the ultimate reason for such a correlation is the presence of
valleys in the random potential. All the wave functions are the
solutions to the Schroedinger equation in the {\it same} random
potential. So, the peaks know where they predominantly want to be
situated. Yet, the criticality matters, as for localized states the
similar correlation is absent, perhaps because there are too many
possibilities to arrange a localized state of small localization
radius. This is the reason why Eq.(\ref{Chalk}) can be formally
extended to ideal metal by setting $d_{2}=d$ but it cannot be
extended to hard insulator which formally corresponds to $d_{2}=0$.
\section{The minimal model in real and in Fock space}
As has been already mentioned in Introduction, in the situation when
Coulomb interaction can be neglected the minimal model consists of
the single-particle Hamiltonian with disorder and the local
attractive interaction \cite{Trivedi}:
\begin{equation}
\label{min} \hat{H}=\sum_{{\bf r},{\bf r'},\alpha} \varepsilon_{{\bf
r},{\bf r'}}\,\Psi^{\dag}_{\alpha}({\bf r})\Psi_{\alpha}({\bf
r'})-U\sum_{{\bf r}}\Psi^{\dag}_{\uparrow}({\bf
r})\Psi^{\dag}_{\downarrow}({\bf r})\Psi_{\downarrow}({\bf
r})\Psi_{\uparrow}({\bf r}).
\end{equation}
In the case where disorder is strong it is very useful to switch
from the coordinate representation to the representation of the
exact single-particle states where all effect of disorder is taken
into account \cite{MaLee,KapKot}. This is done as usual by
substitution $\Psi_{\alpha}({\bf r})=\sum_{n}\psi_{n}({\bf
r})\,c_{n,\alpha}$ of the $\Psi$-operator in terms of the exact
single-particle wavefunctions $\psi_{n}({\bf r})$ and the
creation/annihilation operators  $c^{\dag}_{n,\alpha}$ and
$c_{n,\alpha}$ of electrons in  the corresponding state $n$ with the
spin $\alpha$. As the result the single-particle part of the
Hamiltonian becomes diagonal but the interaction part produces a
variety of terms
\begin{equation}
\label{Ham-Fock}
\hat{H}=\sum_{n,\alpha}E_{n}\,c^{\dag}_{n,\alpha}c_{n,\alpha}
-U\sum_{n,k,m,l}\,M_{nkml}c^{\dag}_{n,\uparrow}
c^{\dag}_{k,\downarrow}c_{m,\downarrow}c_{l,\uparrow},
\end{equation}
where
\begin{equation}
\label{matr-elem} M_{nkml}=\int d^{d}{\bf r}\,\psi^{*}_{n}({\bf r})
\psi^{*}_{k}({\bf r})\psi_{m}({\bf r})\psi_{l}({\bf r}).
\end{equation}
These terms may be divided into two major groups. The first one
consists of the terms where all four indices are the same $n=k=m=l$
or there are two pairs of equal indices. The examples are
$M_{n}\equiv M_{nnnn}$ and $M_{nm}\equiv M_{nnmm}$, as well as
$\bar{M}_{nm}\equiv M_{nmmn}$. They do not change the parity of
occupation of a single particle state but can only move a singlet
pair from state $n$ to state $m$ ($M_{nnmm}$) or to flip the spin of
the singly occupied state ($M_{nmnm}$). All the other terms may
convert the singly-occupied state into the empty or doubly occupied
states and vise versa.

In the region of localized single-particle states the largest matrix
element is $M_{n}$, all the other matrix elements are smaller
because different single-particle states rarely overlap. When the
matrix element $-UM_{n}$ is large enough, it favors a {\it local
pair formation} by producing a gap between the {\it even} many-body
states in which all the single-particle states are occupied by a
singlet pair or empty, and all other ({\it odd}) states. In this
case one may neglect the presence of odd states whatsoever and
introduce bosonic operators
$b_{n}^{\dagger}=c^{\dagger}_{n\uparrow}c^{\dagger}_{n\downarrow}$.
One can easily check that $(b^{\dagger}_{n})^{2}=0$, so that we
introduced a {\it hard-core} bosons. In addition to that, for even
states $b^{\dag}_{n}b_{n}=c^{\dag}_{n\uparrow}c_{n\uparrow}=
c^{\dag}_{n\downarrow}c_{n\downarrow}$.

The {\it minimal model in the Fock space} is obtained by neglecting
of all terms except for those proportional to $M_{n}$ and $M_{nm}$.
The corresponding Hamiltonian takes the form:
\begin{equation}
\label{min-Fock}
\hat{H}=\sum_{n}2E_{n}\,b^{\dag}_{n}b_{n}-U\sum_{n,m}M_{nm}\,b^{\dag}_{n}b_{m}+
h.c.
\end{equation}
It describes physics of local pairs with hard-core interaction
attached to single-particle states $n$ with random energies $2E_{n}$
and hopping due to the attractive interaction of original electrons.
The term proportional to $M_{n}$ is included into the chemical
potential. Obviously, the hopping favors establishing a delocalized
phase which at low enough temperature must be superfluid. Disorder,
in contrast, is trying to localize the pairs.

Eq.(\ref{min-Fock}) admits also a spin representation in which the
hard-core nature of bosons is automatically included. To this end
one introduces the spin-$1/2$ operators
$S^{+}_{n}=b^{\dag}_{n}=c^{\dagger}_{n\uparrow}c^{\dagger}_{n\downarrow}$
and
$S^{z}_{n}=b^{\dag}_{n}b_{n}-\frac{1}{2}=\frac{1}{2}\,[c^{\dag}_{n\uparrow}c_{n\uparrow}+
c^{\dag}_{n\downarrow}c_{n\downarrow}-1]$. Then Eq.(\ref{min-Fock})
takes the form:
\begin{equation}
\label{spin-Fock}
\hat{H}=\sum_{n}2E_{n}\,S^{z}_{n}-U\sum_{n,m}M_{nm}\,S^{+}_{n}S^{-}_{m}+h.c.
\end{equation}
In this language the Ising order corresponds to localized pairs and
the $XY$ order corresponds to a superfluid phase.

So far we neglected the term proportional to $M_{nmmn}$ which is
also operating in the even sector of the Hilbert space. It can be
easily taken into account by adding the following term to the
Hamiltonian:
\begin{equation}
\label{z-z}
-U\sum_{n,m}\bar{M}_{nm}\,S^{z}_{n}S^{z}_{m}=-U\sum_{n,m}
\bar{M}_{nm}\,b^{\dag}_{n}b_{n}b^{\dag}_{m}b_{m}+UV^{-1}\sum_{n}b^{\dag}_{n}b_{n}.
\end{equation}
This term describes interaction (attraction) of bosons at different
points in the Fock space. It favors phase separation with fixed pair
occupation number $1$ or $0$ and thus disfavors establishing a
coherent superfluid phase where the occupation number strongly
fluctuates. This has an effect of reducing the
superconducting/superfluid transition temperature. Such a
suppression of $T_{c}$ by a factor of order 1 \cite{FIKC} does not
change, however, the main conclusions of this paper.

Among other omitted terms the most significant are the following:
\begin{equation}
\label{Hybr} -U\sum_{n,m,l}
M_{nnml}\,b^{\dag}_{n}c_{m\downarrow}c_{l\uparrow} +h.c.
\end{equation}
which describe dissociation and creation of pairs out of singly
occupied single-particle states. They can be safely neglected in the
region of sufficiently strongly localized single-particle states but
have to be taken into account in the region of critical and
sub-critical states where they determine the Ginzburg number ${\rm
Gi}$.
\section{Mean field approximation in the Fock space}
The minimal model Eq.(\ref{spin-Fock}) is a good starting point not
only in the region of localized single-particle states ({\it the
"pseudo-gap" region}) but also in the critical and sub-critical
region, although in this case its derivation from the minimal model
in real space Eq.(\ref{min}) is not justified by a small parameter.
Indeed, the standard mean-field treatment of the spin-model
Eq.(\ref{spin-Fock}) yields the following equation for the critical
temperature \cite{FIKY}:
\begin{equation}
\label{MF-cont-Tc} \Delta_{E}=\frac{\lambda}{2}\int
dE'\,\Delta_{E'}\,\frac{\tanh\left(
\frac{E'}{T}\right)}{E'}\,M_{E-E'},
\end{equation}
where $M_{\omega}$ is given by Eq.(\ref{Chalk}) and
$\lambda=U\nu_{0}$ is the dimensionless attraction constant.

At a weak disorder
 $M_{\omega}=1$ for $|\omega|<\omega_{D}$ and zero otherwise, and one
can immediately recognize the standard BCS equation for $T_{c}$,
albeit with $T$ instead of $2T$. This is the price of restriction to
the odd sector of the Hilbert space. In case when there is no gap
between the many-body states of the even and odd sectors, the latter
is thermodynamically relevant even if dynamically both sectors are
totally decoupled. We will show, however, that the accuracy of the
mean-field approximation on the metallic side close to the Anderson
localization transition is up to a constant pre-factor of order 1.
With this uncertainty the difference by a factor of 2 discussed
above is beyond the accuracy of the mean-field approach.

Let us apply the mean-field equation (\ref{MF-cont-Tc}) to compute
the $T_{c}$ in the critical region. Plugging Eq.(\ref{Chalk}) in
Eq.(\ref{MF-cont-Tc}) and making a rescaling $E\rightarrow
T_{c}\,\varepsilon$, $E'\rightarrow T_{c}\,\varepsilon'$ one reduces
the problem of $T_{c}$ to that of the eigenvalue
$\bar{\lambda}^{-1}\equiv\lambda^{-1}\,(T_{c}/E_{0})^{\mu}\sim 1$ of
the dimensionless linear integral operator. Then we immediately
conclude that:
\begin{equation}
\label{crit-Tc} T_{c}\sim
E_{0}\,\lambda^{\frac{1}{\mu}},\;\;\;\;\mu=1-\frac{d_{2}}{d}.
\end{equation}
\section{Ginzburg number and the accuracy of the mean-field approximation}
As is well known, the thermodynamic fluctuations of phase and
fluctuation of the local $T_{c}$ due to disorder restrict the region
of validity of the mean-field approximation. To take into account
correctly the first effect in the region of {\it extended}
single-particle states one needs to include dissociation processes
Eq.(\ref{Hybr}). Without such processes, for the minimum model in
the Fock space Eq.(\ref{spin-Fock}), the mean-field approximation is
exact in this region, as the matrix element $M_{nm}$ couples to an
infinite number of states in the thermodynamic limit. When
dissociation processes are properly accounted for \cite{FIKC} one
arrives at a remarkable result:
\begin{equation}
\label{Gi} {\rm Gi}\equiv \frac{\Delta T_{c}}{T_{c}}\sim 1,
\end{equation}
where $\Delta T_{c}$ is the "fluctuation region". The effects of
multifractality cancel out in the ${\rm Gi}$ number, and it appears
to be a universal number of order $1$  close to the Anderson
localization transition \cite{KapKot}.

In the region of sufficiently strongly localized single-particle
states, the ${\rm Gi}$ number is mainly controlled by the effective
coordination number $K$ of states coupled by the matrix element
$M_{nm}$ in the minimal spin model Eq.(\ref{spin-Fock}). As one goes
away from the Anderson transition, this number increases. One can
show \cite{FIKC} that in this region:
\begin{equation}
\label{Gi-ins} {\rm Gi}={\rm Gi_{1}}+\frac{{\rm Gi_{2}}}{K^{2}},
\end{equation}
where ${\rm Gi_{1}}$ and ${\rm Gi_{2}}$ are universal numbers of
order 1. This effect of decreasing of coordination number with
increasing disorder invalidates the mean-field approximation at
sufficiently strong disorder and finally leads to the superconductor
to insulator transition.

Note however, that in the vicinity of the Anderson transition
Eq.(\ref{Gi}) holds true. Given the relationship between the
Ginzburg number and the energy scale $J$ associated with the phase
rigidity:
\begin{equation}
\label{Gi-J} J=\frac{T_{c}^{(MF)}}{{\rm Gi}^{\frac{1}{3}}}
\end{equation}
one concludes that the scale $J$  and the mean-field transition
temperature $T_{c}^{(MF)}$ given by Eq.(\ref{crit-Tc}) differ by a
universal factor of order 1. This implies that the phase
fluctuations may reduce the true superconducting transition
temperature by at most a universal factor of order 1 compared to the
mean-field result Eq.(\ref{crit-Tc}).

\section{Virial expansion method}
In order to describe the behavior of transition temperature in the
region of localized single-particle states where the mean-field
approach breaks down, we exploit \cite{FIKC} the idea of virial
expansion applied to the spin-model Eq.(\ref{spin-Fock}). To this
end we analytically express the Cooper susceptibility $\chi_{C}(T)$
(the response of the spin $S^{+}$ to the infinitesimal perturbation
$\Delta\,S^{+}$) as a sum of contributions of clusters of $M$
coupled spins:
\begin{equation}
\label{virial} \chi_{C}(T)=\sum_{M+1}^{\infty} \chi_{M}(T),
\end{equation}
The true transition temperature corresponds to temperature $T_{c}$
when this series becomes divergent. According to D'Alembert
criterion this happens when:
\begin{equation}
\label{Dalembert} \lim_{M\rightarrow\infty}
\frac{\chi_{M+1}(T_{c})}{\chi_{M}(T_{c})}=1.
\end{equation}
In view of rapid increase of complexity of the analytical
expressions for $\chi_{M}$ with increasing $M$, we truncated the
series Eq.(\ref{virial}) and used the operative definition of
$T_{c}$ in terms of small-cluster contributions $T_{c}^{(1)}$ and
$T_{c}^{(2)}$:
\begin{equation}
\label{oper}
\chi_{1}(T_{c}^{(1)})=\chi_{2}(T_{c}^{(1)}),\;\;\;\;
\chi_{2}(T_{c}^{(2)})=\chi_{3}(T_{c}^{(2)}).
\end{equation}
By evaluating $\chi_{1}$, $\chi_{2}$ and $\chi_{3}$ taking the
energies $E_{n}$ and matrix elements $M_{nm}$ from exact
diagonalization of the Anderson model on a 3D lattice and solving
Eq.(\ref{oper}) numerically we were able to find statistics of
$T_{c}^{(1)}$ and $T_{c}^{(2)}$ at different positions of Fermi
level relative to the mobility edge. The average $T_{c}^{(1)}$ and
$T_{c}^{(2)}$ appear to be in good agreement with each other which
gave us a reasonable confidence in the convergence of the procedure.
\begin{figure}[h]
\includegraphics[width=12cm, height=10cm,angle=0]{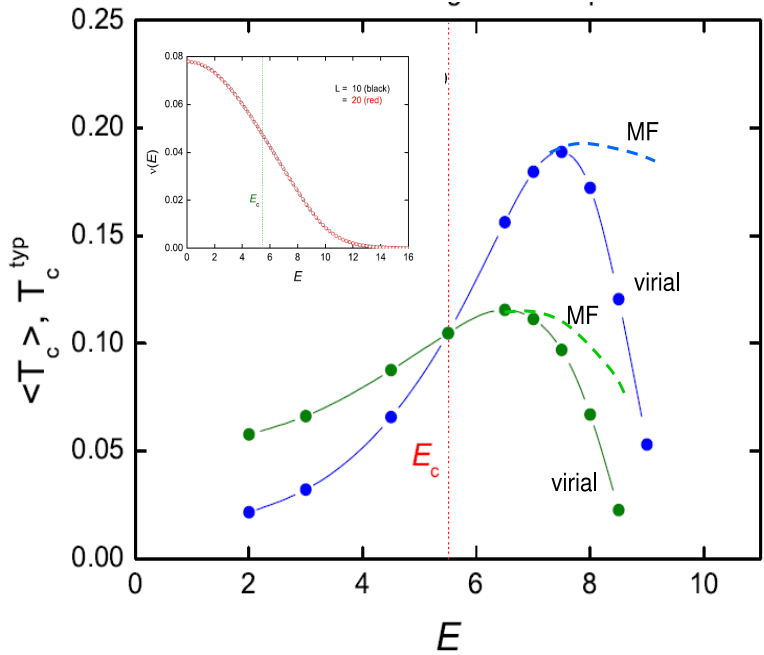}
\caption{(color online)Transition temperature  as a function of the
Fermi energy position relative the mobility edge $E_{c}$ for the
fixed dimensionless constant $\lambda=0.08$ (blue circles) and for
the fixed attractive interaction $U=1.7 t$(green circles), where $t$
is the hopping integral of the Anderson model. In the insert: energy
dependence of the mean density of states. Large energy corresponds
to region of localized single-particle states. Inside this region
the mean-field (MF) transition temperature is substantially larger
than that ($T_{c}\equiv\langle T_{c}^{(2)}\rangle$) found from the
truncated virial expansion series.   The Cooper susceptibilities
$\chi_{2}(T)$ and $\chi_{3}(T)$ were calculated numerically using
exact eigenfunctions and eigenvalues of the Anderson model with
strength of disorder $W=4$, size of system $L=20^{3}$ and the Debye
frequency $\omega_{0}=0.5$.}
\end{figure}
In Fig.3, the results of virial expansion calculation for the
average $\langle T_{c}^{(2)} \rangle$ are presented as a function of
the Fermi level position. It is remarkable that the truncated virial
expansion method is in good agreement with the results of mean-field
calculations in the region of extended and weakly localized states
and gives substantially smaller transition temperature for stronger
localized single-particle states. This proves that the virial
expansion method, even with only 3 terms retained in the series
Eq.(\ref{Dalembert}), captures the decrease of the effective
coordination number $K$ and is not equivalent to the mean-field
approximation.
\section{Enhancement of $T_{c}$ near the Anderson transition.}
Equation (\ref{crit-Tc}) shows that when the Fermi energy is at the
mobility edge $E=E_{c}$ the critical temperature behaves as a power
law of the dimensionless attraction constant $\lambda$. Numerical
simulations on the 3D Anderson model of localization give (see Ref.
\cite{FIKC} and references therein) the value of the fractal
dimension $d_{2}=1.29\pm 0.1$. Thus the exponent in the power law
dependence of $T_{c}$ on $\lambda$ is close to $1/\mu\approx 1.8$.
For small $\lambda$ one always has $\lambda^{1.8}\gg {\rm
exp}(1/\lambda)$. In addition to that the characteristic energy
$E_{0}$ is of electronic nature and it is typically higher than the
Debye frequency $\omega_{0}$. Therefore we conclude that at small
$\lambda$ the transition temperature given by Eq.(\ref{crit-Tc}) is
much higher than the BCS transition temperature $T_{c}\sim
\omega_{0}\,{\rm exp}(-1/\lambda)$. This enhancement of $T_{c}$ is
also obtained by a virial expansion method which is completely
independent of the mean-field approach. It can be easily traced back
to the enhancement of the matrix element $M_{nm}$ shown in Fig.1.

In order to figure out how general is the effect of enhancement of
$T_{c}$ at negligible Coulomb interaction, let us consider the case
of weak multifractality $1-d_{2}/d \ll 1$. This case is relevant for
the 2D metal \cite{AKL,EF}, where
\begin{equation}
\label{frac} 1-\frac{d_{2}}{2}=\frac{1}{g},
\;\;\;\;g=2\pi^{2}\nu_{0}D.
\end{equation}
Expanding the matrix element $Eq.(\ref{Chalk})$ in the mean-field
equation Eq.(\ref{MF-cont-Tc}) up to the leading correction in $1/g$
$M_{\omega}\approx 1+\frac{1}{g}\,\ln(E_{0}/\omega)$ we obtain the
$1/g$ correction to the transition temperature:
\begin{equation}
\label{corr} \frac{\delta
T_{c}}{T_{c}}=\frac{1}{2g}\,\ln\left(\frac{\omega_{0}}{T_{c}}\right)\,\ln\left(\frac{E_{0}^{2}}
{\omega_{0}T_{c}}\right)\sim
\frac{1}{g}\,\ln^{2}\left(\frac{\omega_{0}}{T_{c}}\right).
\end{equation}
We obtained the correction of the same order as in Ref.\cite{MayFuk}
given by Eq.(\ref{MayFuk-attrac}) thus proving that our analysis at
$\lambda\ll 1$ qualitatively applies also to the case of {\it weak
multifractality} in 2D metals. Then Eq.(\ref{MayFuk-Coulomb}) tells
us that the really crucial assumption for enhancement of $T_{c}$ by
disorder is the suppression of Coulomb interaction. Recently
\cite{BurMir} this statement was confirmed by the RG analysis of the
Finkelstein-like nonlinear sigma-model in $d=2$ and  $d=2+\epsilon$
where only the short-range interactions were taken into account.

There are certain systems \cite{Zr,Iwasa} where one may suspect such
a suppression. Moreover, the enhancement of $T_{c}$ just before the
onset of the insulating behavior was observed in one of them with
the dependence of $T_{c}$ on doping highly reminiscent of the Fig.3.
More investigations are needed to confirm or reject this conjecture.
In any case, the search for systems with suppressed Coulomb
interaction (except for obvious case of cold neutral fermionic atoms
with attractive interaction) is a challenging task which may open a
door into a wonderful world without Coulomb interaction.

\end{document}